\documentclass[10pt,twocolumn]{article}

\usepackage{multicol}
\usepackage{graphicx}
\usepackage{amsmath}
\usepackage{amssymb}
\usepackage{lipsum}
\usepackage{color}
\usepackage{siunitx}
\usepackage{authblk}
\usepackage[margin=2cm]{geometry}

\title{Tunable transport of drops on a vibrating inclined fiber}
\author[1]{Alison Bick}
\author[1]{Fran\c{c}ois Boulogne}
\author[1]{Alban Sauret}
\author[1]{Howard A. Stone}
\affil[1]{ Department of Mechanical and Aerospace Engineering, Princeton University, Princeton, NJ 08544}
\date{\today}
\begin{document}


\twocolumn[
    \begin{@twocolumnfalse}
        \maketitle
        \begin{abstract}
            Transport of liquid drops in fibrous media occurs in various engineering systems such as fog harvesting or cleaning of textiles.
    The ability to tune or to control liquid movement can increase the system efficiency and enable new engineering applications.
    In this Letter, we experimentally investigate how partially wetting drops on a single fiber can be manipulated by vibrating the fiber.
    We show that a sliding motion along the fiber or a dripping of the drop can be triggered by standing waves.
    We identify the conditions on the drop volume, the fiber tilt angle and the amplitude and frequency of oscillations to observe these different behaviors.
    Finally, we experimentally illustrate that vibrations can be used to control the transport and the collection of water drops along a fiber using a combination of the sliding and dripping transitions.
        \end{abstract}
    \end{@twocolumnfalse}
]

The capture of liquid on a network of fibers is encountered in various
practical applications. For instance, fog and dew harvesting has been shown
to be a promising sustainable and environmentally friendly method to
collect water in arid regions. \cite{Olivier2004} Fog is composed of many small water
droplets that are transported by the wind and can impact and be collected
by a net. \cite{Dressaire2015} However, the reentrainment of water in the air and the
clogging of the fiber network then represents the major limitations to the
improvement of the net's efficiency.\cite{Park2013} Indeed, when liquid collects on an array of fibers,
the drops can attain a stable configuration between the fibers,\cite{Mullins2004,Sauret2014,EPJE2015}
preventing liquid collection and decreasing the efficiency of such a net.
To optimize the design and the efficiency of fog nets, three-dimensional hierarchical
structures have been proposed,\cite{Andrews2011,Azad2015} which can be coupled
with chemical surface properties of the material to alter the contact angle
hysteresis.\cite{Park2013} Thus, the transport of water on the fog nets
appears to be crucial to improve their efficiency.

Whereas a natural passive method for transport consists of using gravity to slide
drops along a fiber,\cite{Duprat2009b,Gilet2010,Boulogne2012} active
methods could be used to collect water more efficiently. Indeed, a large
variety of active methods have been proposed to induce spontaneous drop
motion in various geometries. For instance, we can mention capillary
pressure driven propulsion due to asymmetric drops \cite{Haefner2015} or
gradient wettability, \cite{Lorenceau2004,Ju2013} Leidenfrost drops on an anisotropic
rough substrate,\cite{Lagubeau2011} drop interactions due to
evaporation-induced surface tension gradients \cite{Cira2015} or
thermal gradients.\cite{Brochard1989,Brzoska1993} Mechanical forcing could also be considered to trigger drop motion. Indeed, Noblin \textit{et al.}
have shown that drop motion can be induced on an
oscillating horizontal plane,\cite{Noblin2004} which highlighted transitions due to
pinned/mobile contact lines. Naturally, drops can slide down a
plane \cite{Legrand2005,Snoeijer2007} but the opposite motion has been
reported experimentally and theoretically if an inclined plane is oscillated with particular asymmetries and ranges of frequencies, amplitudes and directions.\cite{Daniel2005,Dong2006,Brunet2007,Brunet2009,Benilov2011}
Also, drop motion has been induced with combination of horizontal and vertical oscllations\cite{Noblin2009} or with micro-structured texture ratchets\cite{Duncombe2012}.

In this Letter, we investigate experimentally the motion of a partially wetting drop on a different geometry, a
deformable fiber, in which oscillations are induced at one of the fiber's ends.
We first identify the conditions necessary to observe static, sliding or dripping
behavior on a static inclined fiber. Then, we show that these conditions can
be tuned by the application of a standing wave on the fiber. Finally, we
illustrate how the sliding motion induced by oscillations can be used for
water collection.

\begin{figure}
    \begin{center}
    \includegraphics[width=.85\linewidth]{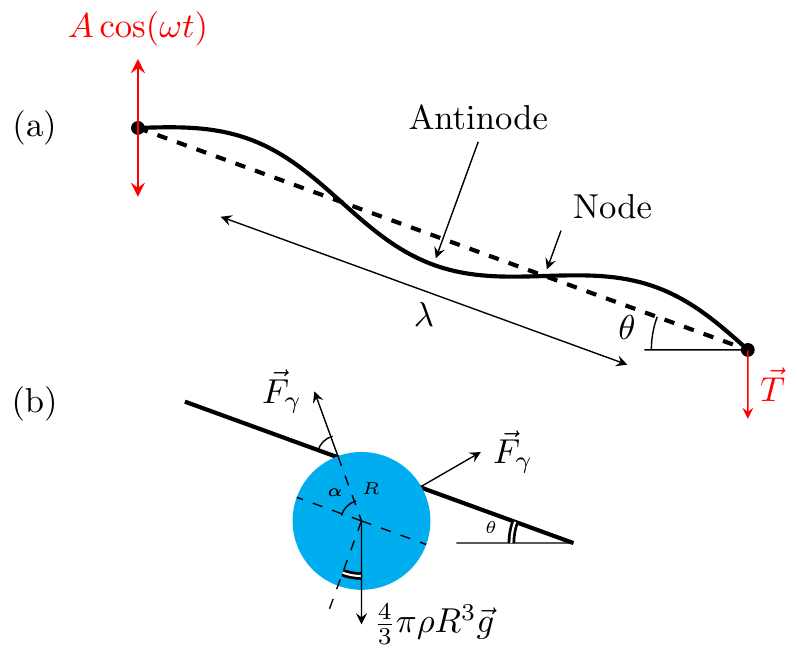}
    \caption{
        (a) Schematic of the experimental setup.
        A nylon fiber of radius $b$ is inclined from the horizontal by an angle $\theta$ and oscillated at the upper end; a fixed tension $T$ is applied at the lower end.
        The frequency $\omega$ of the oscillations of amplitude $A$ correspond to standing waves of wavelength $\lambda$.
        (b) Schematic of a drop on a tilted fiber.
    Here, $\alpha$ is the angle from the center of the drop to the fiber, $R$ is the radius of the drop, $g$ is the gravitational acceleration, and $F_\gamma = 2 \pi b \gamma \sin \alpha$ is the capillary force.}
    \label{fig:setup}
    \end{center}
\end{figure}

The experimental setup,\cite{Melde1860} depicted in Fig. \ref{fig:setup}, consists of a tilted nylon fiber (diameter $2b=0.35$ mm, $18.3$ g/m weight per unit length) attached on the upper end to a mechanical vibrator (LDS 319024-3), which is controlled by a sinusoidal wave generator (Stanford Research System Model DS345) and an amplifier (LDS PA100E).
A constant tension, $T=0.5$ N, is applied at the lower end of the nylon fiber.
The oscillation frequency, typically $f=\omega /(2\pi)\approx 100$ Hz, is chosen to obtain standing waves of wavelength $\lambda\in[20,40]$ cm and an amplitude of oscillation $A\in[0,4]$ mm.
Drops of water with blue dye (McCormick Food Color) are dispensed on the fiber with a micropipette.
In the following, this solution is referred to as water and the volume is in the range $[2, 18]$ $\mu$L.
The density of the water with dye is $\rho=1000$ kg/m$^3$ and the surface tension $\gamma=71$ mN/m, which is measured with a pendant drop technique.
Images are recorded using a high-speed camera (Phantom V9.1 with Nikon 85 mm Macro lens) typically operating at 1000 frames per second.
The lighting is provided by an LED panel (Phlox, 10 cm $\times$ 10 cm) placed behind the fiber.

\begin{figure}
    \centering
    \includegraphics[width=1\linewidth]{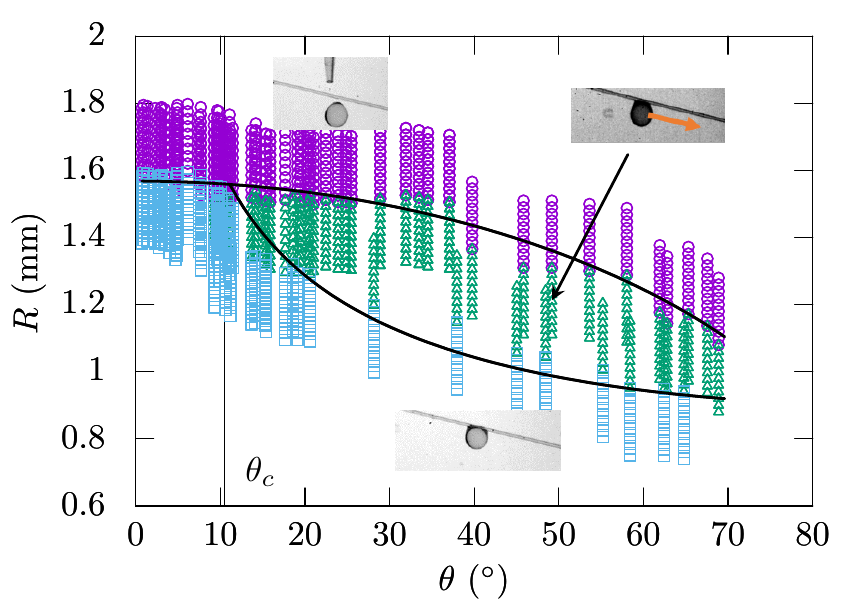}
    \caption{
        Phase diagram of drop behavior on a static fiber as a function of the drop radius $R$ and the fiber inclination angle $\theta$.
        Three distinct domains are observed: static ($\square$), sliding ($\bigtriangleup$) and dripping ($\bigcirc$).
        The upper and lower solid lines represent equations (\ref{eqn:RM}) and (\ref{eqn:Rs}) respectively.
        In equation (\ref{eqn:Rs}), we used $F_f^{max} = 3\times 10^{-5}$ N.
        The angle $\theta_c = 11^\circ$ denotes the critical angle above which the three domains can be observed.
    }
    \label{fig:phaseDiagram}
\end{figure}

For a static fiber, $A=0$, we report in Fig. \ref{fig:phaseDiagram} the drop behavior as a function of its radius $R$ and the fiber inclination $\theta$.
Three regimes are observed: static, sliding and dripping.
For the sliding behavior, since the drop is partially wetting the fiber, no coating is observed behind the drop.
To describe the dripping transition, we consider the force balance between gravity and capillary forces\cite{Lorenceau2004}
\begin{equation}
    2F_\gamma \cos\theta = \frac{4}{3} \pi \rho g R^3,
\end{equation}
where we used Tate's law\cite{Tate1864} for the capillary force $F_\gamma = 2 \pi b \gamma \sin\alpha$.
Assuming that dripping occurs for $\alpha = {\pi}/{2}$, the critical radius of the drop is
\begin{equation}
    R_d=\left(3 b {\ell_c}^2 \cos \theta   \right)^{1/3},  \label{eqn:RM}
\end{equation}
where the capillary length is $\ell_c= \sqrt{{\gamma}/({\rho g})}$.
As the tilt angle of the fiber is increased, a sliding regime between the static and dripping regimes is observed for $\theta>\theta_c$ (Fig. \ref{fig:phaseDiagram}).
To model this sliding transition, we introduce a friction force $F_f$ to describe phenomenologically the pinning force of the contact line, which balances the drop weight in the static domain of the phase diagram.
Note that this friction force depends in particular on the liquid wetting properties and on the roughness of the material.
This friction force equilibrates the tangential component of the drop weight
\begin{equation}
    F_f =  \frac{4}{3} \pi \rho g \sin \theta R^3.\label{eqn:frictionforce}
\end{equation}
Thus, the critical radius for the sliding transition is
\begin{equation}
    R_s=\left(\frac{3 F_f^{max}}{4 \pi \rho g \sin\theta}\right)^{1/3} ,  \label{eqn:Rs}
\end{equation}
where $F_f^{max}$ is the maximum pinning force, which is a fitting parameter.
We assume that the drop remains spherical and $F_f^{max}$ is independent of $\theta$.
Equations (\ref{eqn:RM}) and (\ref{eqn:Rs}) are plotted in Fig. \ref{fig:phaseDiagram} and show fairly good agreement for describing the different transitions over a range of tilt angles $\theta \in [0, 70^{\circ}]$.

\begin{figure}
    \centering
    \includegraphics[scale=1]{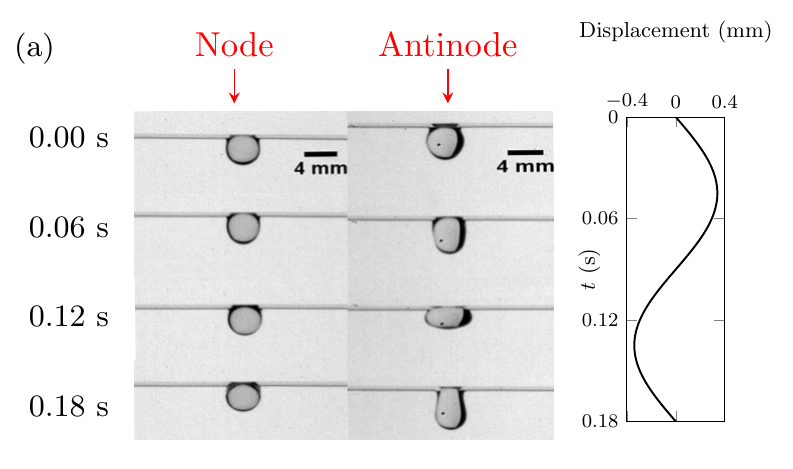}\\
    \includegraphics[width=1\linewidth]{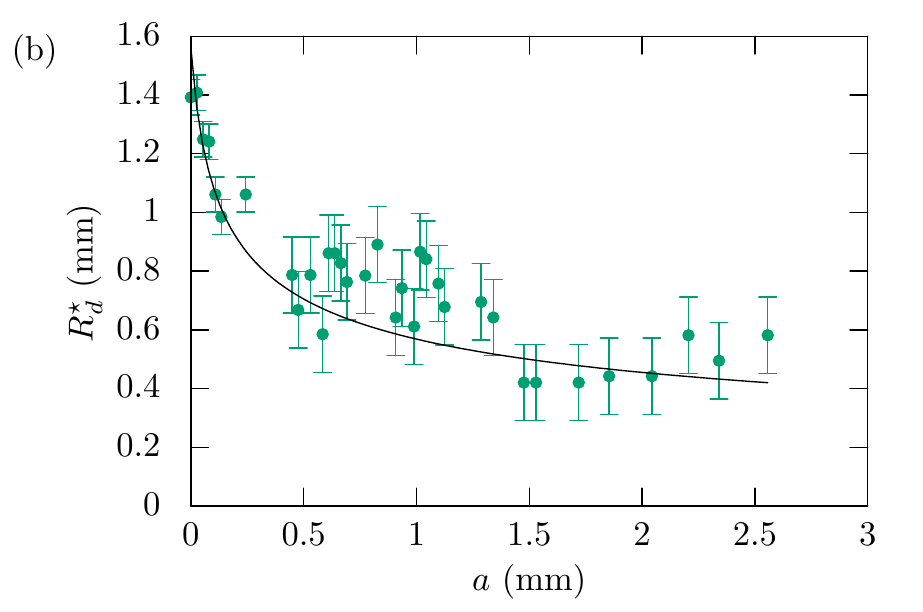}
    \caption{Drop motion on a horizontal vibrating fiber. (a) Time-lapse image of a water drop placed on a node and an antinode.  The plot shows the displacement imposed at the upper end of the fiber. The drop volume is 5 $\mu$L, the oscillation amplitude and frequency are 0.34 mm and $f=140$ Hz, respectively. (b) Critical radius after which the drop detaches from the fiber versus the local amplitude $a$ of the fiber. The solid line is equation (\ref{eqn:RM*}).
    }
    \label{fig:vibHorFiber}
\end{figure}

We now examine the drop dynamics on a horizontal vibrating fiber, $\theta=0^{\rm o}$.
The drop behavior depends on its location on the fiber.
For a drop located at the node, no measurable deformation of the drop is observed in the range of frequencies and amplitude used in our experiments  as illustrated in Fig. \ref{fig:vibHorFiber}a.
On the contrary, for a drop placed at an antinode, the drop exhibits significant deformations.
We also notice that the drop remains at the same position on the fiber and the drop deformation is out of phase with respect to the fiber motion.
As the amplitude of oscillation is increased, the deformation of the drop increases until a critical amplitude triggers a dripping transition, which depends on the drop volume.
Because of the acceleration of the fiber, the maximum apparent gravity is
\begin{equation}
    g^\star = g + \omega^2 a\label{eqn:apparent_gravity}
\end{equation}
where $a$ is the local amplitude of oscillation of the fiber where the drop is positioned.
By redefining the capillary length $\ell_c^\star = \sqrt{{\gamma}/({\rho g^\star})}$ with this apparent gravity, the critical radius of a drop on an oscillating horizontal fiber is
\begin{equation}
    R_d^\star=\left(3b{\ell^\star_c}^2 \right)^{1/3}.  \label{eqn:RM*}
\end{equation}
We report in Fig. \ref{fig:vibHorFiber}b the critical amplitude to observe the dripping transition obtained experimentally for various drop volumes.
The transition is well-captured by the scaling law (\ref{eqn:RM*}).

\begin{center}
\begin{figure}
    \includegraphics[width=0.9\linewidth]{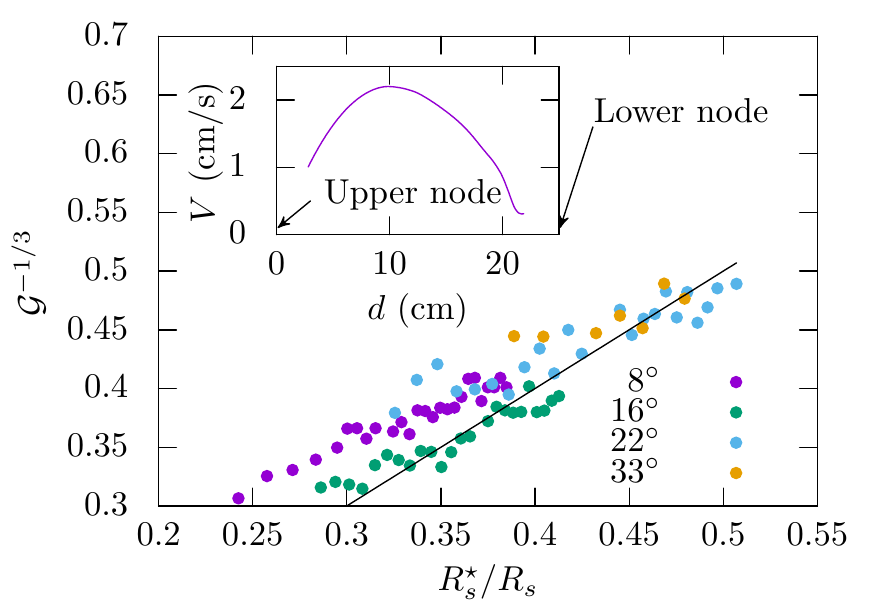}
    \caption{ Dimensionless gravity ${\cal G} = (1 + \omega^2 a/g)$ as a function of the dimensionless critical radius $R_s^\star/R_s$ for sliding for different fiber tilt angles $\theta$. For each radius, we identify the critical oscillation amplitude for sliding. The value $F_f^{max} = 3\times 10^{-5}$ N is used to define $R_s$. The solid line shows ${\cal G}^{-1/3} = R_s^\star/R_s$. The typical errors on ${\cal G}$ are $\pm 0.02$. Inset: Drop velocity $V$ as a function of position $d$ along the fiber from an upper node to a lower node. The frequency is $f=140$ Hz, the tilt  angle $\theta=22^\circ$, and the drop volume is 5.5 $\mu$L.}
    \label{fig:vibTiltFiber}
\end{figure}
\end{center}

Finally, we placed a single drop on a static tilted fiber.
We chose the drop volume and the angle of the fiber to ensure that the drop is static on the fiber (Fig. \ref{fig:phaseDiagram}).
When the fiber is vibrated, we observe that above a critical amplitude, the drop slides down the fiber.
For a drop placed a few centimeters downstream from a node, the drop accelerates while approaching the antinode and then decelerates toward the next node (see inset of Fig. \ref{fig:vibTiltFiber}).
For the conditions of our experiments, the drop stops at a distance of the order of $1$ cm preceding this node.

The critical amplitude to trigger the sliding motion depends on the fiber angle $\theta$ and the drop radius.
As for the dripping transition, we consider that fiber oscillations results in an apparent  gravity given by equation (\ref{eqn:apparent_gravity}).
From equation (\ref{eqn:Rs}), the critical radius of the drop for the sliding transition can be generalized as
\begin{equation}
    R_s^\star= R_s {\cal G}^{-1/3},\label{eqn:slipping}
\end{equation}
where ${\cal G} = (1+\omega^2 a / g)$.

To compare equation (\ref{eqn:slipping}) with the experiments, a drop is dispensed on a tilted fiber and its volume is chosen to be in the static domain (Fig. \ref{fig:phaseDiagram}).
In Fig. \ref{fig:vibTiltFiber}, we report measurements of the critical amplitude to trigger the sliding transition as a function of the drop radius for four different tilt angles.
Equation (\ref{eqn:slipping}) provides reasonable agreement in the range of drop volumes accessible experimentally. Note that the value of $F_f^{max}$ is the same as in the static situation (Fig. \ref{fig:phaseDiagram}).
We also notice that the critical angle $\theta_c$ above which the three behaviors can be observed is reduced.
Indeed, for a tilt angle $\theta = 8^\circ$, the drop slides down the fiber above a critical amplitude (Fig. \ref{fig:vibTiltFiber}).
These results illustrate that the motion of a drop of a specific volume can be triggered by the amplitude of the wave.

\begin{figure}
    \centering
    \includegraphics[width=.95\linewidth]{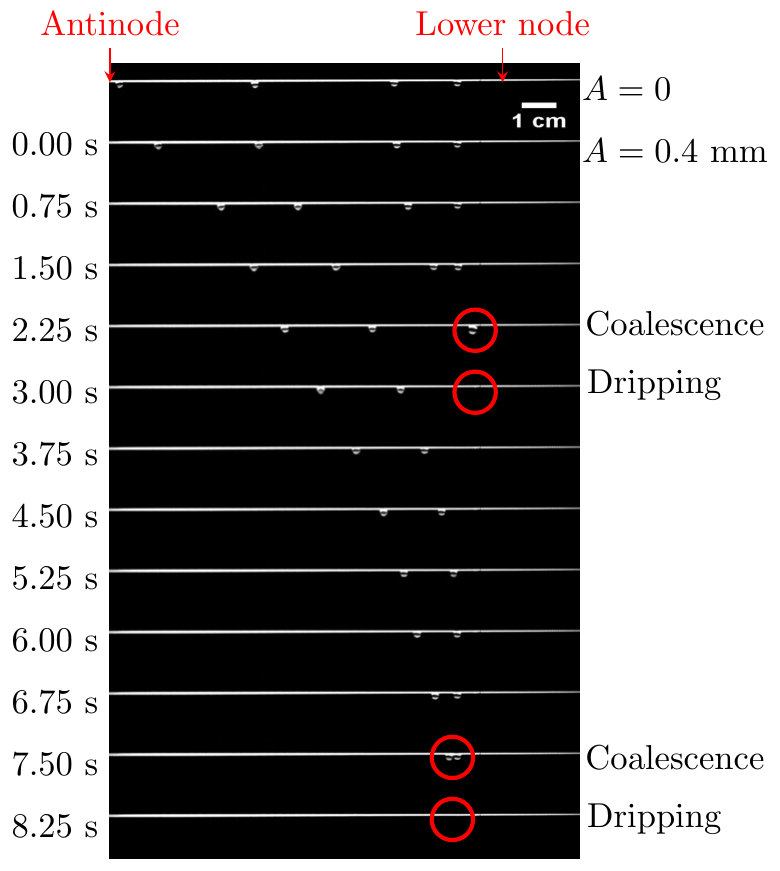}
    \caption[Multiple Drop System placed on vibrating fiber]
    {Illustration of  multiple drop manipulation on an inclined fiber. Initially, evenly spaced drops are placed on the fiber. As vibrations start, the drops slide, coalesce near the node and fall off the fiber. For this experiment $\theta = 16^\circ$, $f=100$ Hz, $A$=0.4 mm and the drop volume is 5 $\mu$L. (Multimedia view)
    }
    \label{fig:multiDropSystem}
\end{figure}

In this Letter, we have shown that on tilted fibers a drop can adopt three distinct behaviors depending on its volume and the fiber angle: static, sliding or dripping.
For static drops, we demonstrated that fiber oscillations can be used to manipulate the drops.
On horizontal fibers, above a critical amplitude and frequency, a drop detaches from the fiber without translational motion.
On tilted fibers, these oscillations can trigger a sliding motion of a drop toward the next lower node where the drop stops.

These observations can be extended to situations involving multiple drops, which are for instance encountered in systems to collect water, such as fog harvesting nets.
In Fig. \ref{fig:multiDropSystem} (Multimedia view), we illustrate the dynamics of four drops placed between two nodes on a fiber subject to standing waves.
Once the oscillations start, drops move along the fiber toward the lower node where their velocity decreases as explained in Fig. \ref{fig:vibTiltFiber}.
Thus, drops coalesce and reach a critical volume above which the drop detaches from the fiber (Fig. \ref{fig:vibHorFiber}).

Using this idea, the drop motion can be controlled and the location of a drop falling off a fiber can be predicted \textit{i.e.} dripping occurs at a node.
So water collection containers can be placed under these locations to increase the amount of water collected.
This water collection method has the potential to improve water collection, in particular for fog harvesting nets.
Future work should consider the importance of the surface microstructure encountered on natural fibers, which may have significant effects on the wetting properties.\cite{Ito2015}

We thank Professor Jay Benziger for useful discussions.
F.B. acknowledges that the research leading to these results received funding from the People Programme (Marie Curie Actions) of the European Union's Seventh Framework Programme (FP7/2007-2013) under REA grant agreement 623541.

%
%

    \end{document}